\documentclass[a4paper,11pt]{article}
\usepackage{jinstpub} 
\usepackage{silence}
\usepackage{lineno}
\usepackage{caption}
\usepackage{subfigure,dcolumn}
\usepackage{multirow}

\title{\boldmath R\&D of cosmic ray detection module with liquid scintillator and wavelength shift fiber}

\author[a]{Jun Zou,}
\author[b,c,1]{Xiangdong Sheng,\note{Corresponding author.}}
\author[d,e,2]{Zhimin Wang,\note{Corresponding author.}}
\author[a,3]{Fengjiao Luo,\note{Corresponding author.}}
\author[a,4]{Bo Zheng,\note{Corresponding author.}}
\author[i]{Cunfeng Feng,}
\author[b,c]{Chao Hou,}
\author[g]{Guang Luo,}
\author[d]{Sibo Wang,}
\author[h]{Peisheng Niu,}
\author[h]{Fang Liu,}
\author[d,e]{Yichen Zheng,}
\author[i]{Dong Liu,}
\author[i]{Ziqi Huang,}
\author[i]{Shulong Ji}

\affiliation[a]{University of South China, Hengyang Hunan 421001, China}
\affiliation[b]{State Key Laboratory of Particle Astrophysics, Institute of High Energy Physics, Chinese Academy of Sciences, Beijing 100049, China}
\affiliation[c]{Tianfu Cosmic Ray Research Center, 610000 Chengdu, Sichuan, China}
\affiliation[d]{Institute of High Energy Physics, Beijing 100049, China}
\affiliation[e]{University of Chinese Academy of Sciences, Beijing 100049, China}
\affiliation[f]{State Key Laboratory of Particle Detection and Electronics, Beijing 100049, China}
\affiliation[g]{Sun Yat-sen University, Guangzhou, China}
\affiliation[h]{North China Electric Power University, Beijing, China}
\affiliation[i]{Institute of Frontier and Interdisciplinary Science, Shandong University, 266237 Qingdao, Shandong, China}

\emailAdd{shengxd@ihep.ac.cn}
\emailAdd{wangzhm@ihep.ac.cn}
\emailAdd{luofengjiao@usc.edu.cn}
\emailAdd{zhengb@usc.edu.cn}

\abstract{
For neutrino physics and rare event searches, background related to cosmic muons poses a notable challenge, and must be identified and rejected. It is also a challenge to control the cost with good performance for a large array of cosmic ray detection.
We proposed a cosmic ray detection module with liquid scintillator and wavelength-shifting fibers for its reasonable cost and performances.
The results from the measurements of a prototype with Muon indicate that the detector's photoelectron response is good. 
The outcomes of this study hold significant potential for applications in cosmic ray observation experiments and underground rare-event detection, providing a viable option for future large-scale observatories.
This work highlights the feasibility of liquid scintillator-based detectors in addressing current and emerging challenges in particle physics and astrophysics.
}

\keywords{Cosmic ray, Liquid scintillator, Wavelength-shifting Fiber}

\arxivnumber{1234.56789} 

\begin{document}
\maketitle
\flushbottom

\section{Introduction}\label{sec:intro}

With the expansion of particle physics and particle astrophysics research topics and experimental scales, there is an increasing demand for large-area, high-efficiency, fast-response, low-cost and modular cosmic ray and electromagnetic particle detectors for large-scale cosmic ray observing arrays. 
The large-scale research, production and application of liquid scintillators (LS) in the Jiangmen Underground Neutrino Observatory (JUNO)\,\cite{JUNO-2022103927} have demonstrated their obvious advantages in terms of cost and other aspects. 
Research on plastic scintillators with fiber modules, conducted as part of the Taishan Antineutrino Observatory (TAO)\,\cite{junocollaboration2020taoconceptualdesignreport,Luo:2023inu,luo_performance_2025} and the Large High Altitude Air Shower Observatory (LHAASO)\,\cite{ZHEN2019457} has yielded significant experience. This experience, combined with advancements in photon-electronic detection technologies (e.g., silicon photomultiplier tubes, SiPM), integrated electronics, and modular power supplies, has enhanced the potential for good and cheap modular electromagnetic particle detectors.
Few previous studies tried different schemes with LS and wavelength shifting (WLS) fiber as in \cite{Zhangyongpeng_2017,Zhangying_2025}.
However, certain challenges and uncertainties remain, and further research is needed to optimize the design and validate the scheme, particularly regarding the detector response to particles of different energies, the ability to exclude background signals, environmental adaptability, and stability under various design scenarios.

Supported by the State Key Laboratory of Particle Detection and Electronics\,\cite{State-Key-Laboratory-of-Particle-Detection-and-Electronics}, and combining the experiences from the JUNO, the Taishan Antineutrino Observatory, and the LHASSO Electromagnetic Particle Detectors (ED)\cite{wang_testing_2021}, the research aims to develop a large-area, high-efficiency, modular, fast-response, and low-cost liquid scintillator-based detectors of cosmic rays/electromagnetic particles in square meter scale, and to form a candidate for a larger-scale cosmic observatory array in the future.
The research results are of great significance and have a wide range of prospects for applications in scenarios such as cosmic observation experiments and underground rare-event observation experiments.

In the following sections, the structure of the module and its performance results are described in detail.

\section{Design \& setup}
\label{sec:design}

\subsection{Detector design \& construction}

\begin{figure}[!ht]
\centering
\subfigure[Detector top view]{\includegraphics[width=0.4\textwidth]{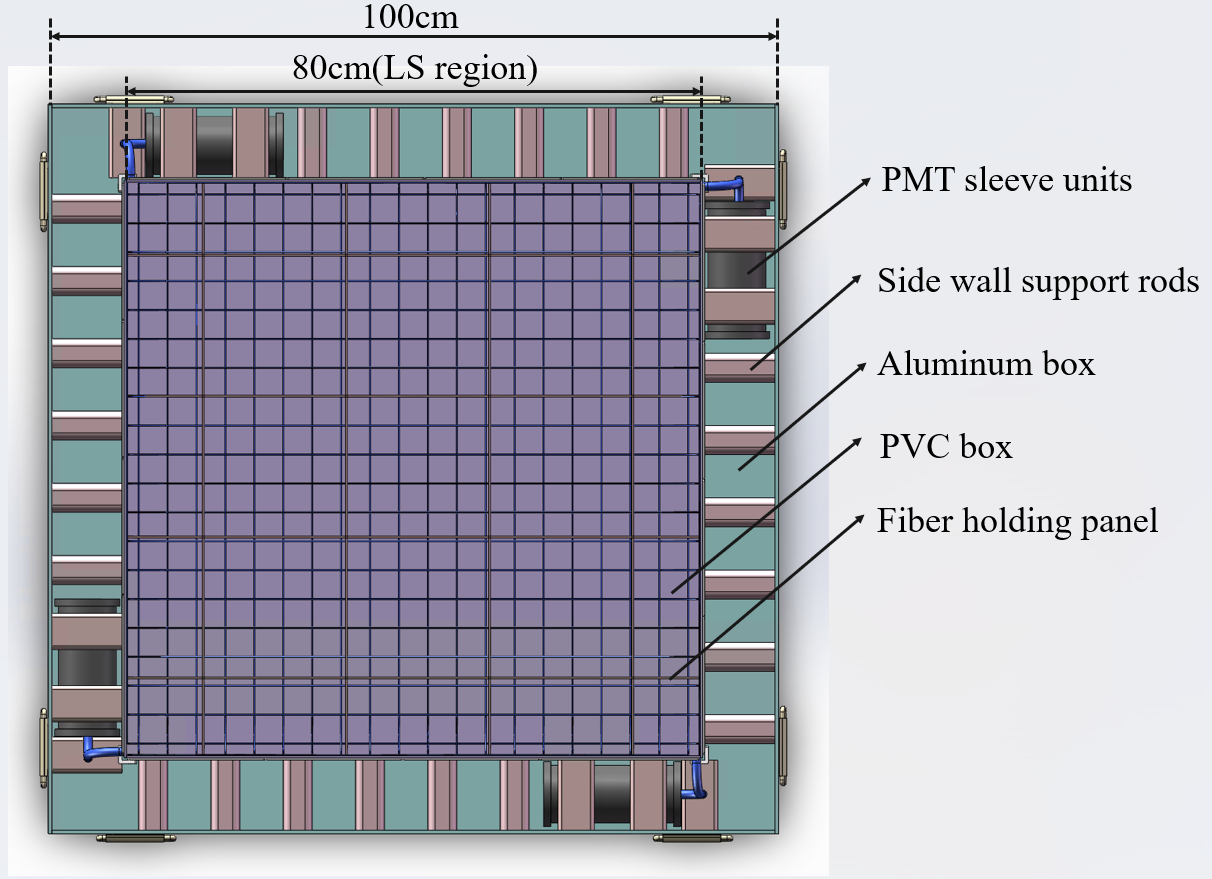}}
\subfigure[Detector side view]{\includegraphics[width=0.42\textwidth]{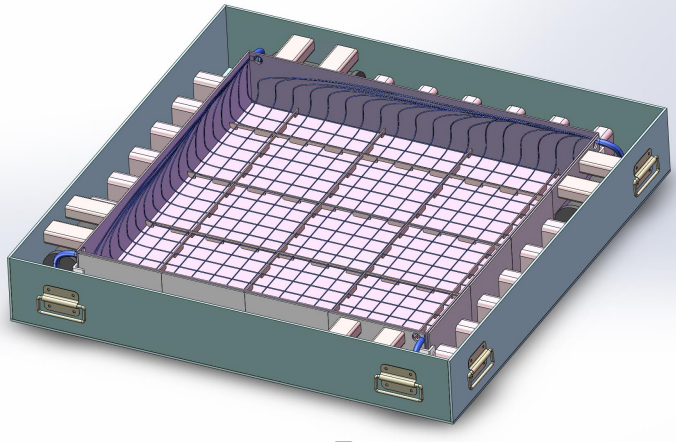}}
\subfigure[Holding panel]{\includegraphics[width=0.4\textwidth]{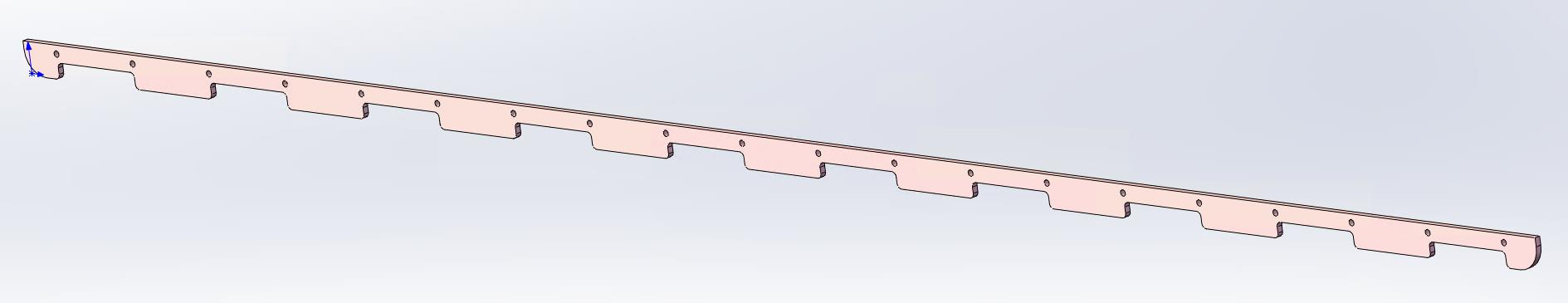}}
\subfigure[PVC corner]{\includegraphics[width=0.4\textwidth]{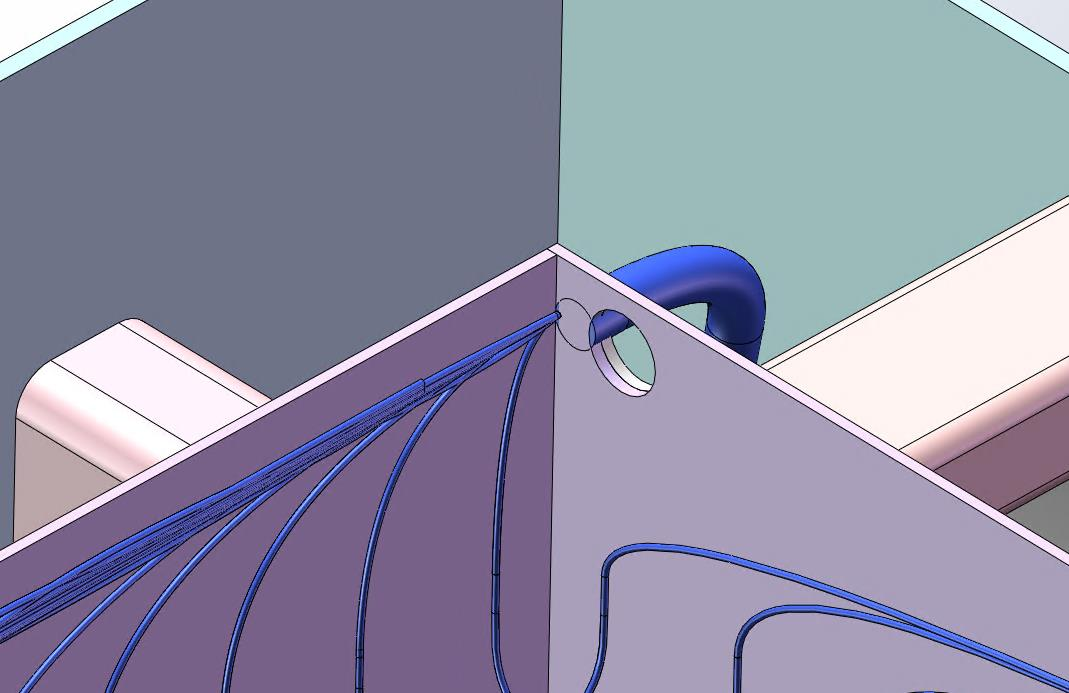}}
\caption{Prototype module}
\label{fig:prototype module}
\end{figure}

A schematic design of the detector module is illustrated in Figure~\ref{fig:prototype module}(a) and \ref{fig:prototype module}(b).
The system consists of an outer aluminum shell box (100\,cm × 100 cm × 15\,cm) that houses a white Polyvinyl chloride (PVC) container (80\,cm × 80\,cm × 12.5\,cm). PVC was chosen as the container material due to its good compatibility with LS and its cost-effectiveness. The PVC container is filled with LS\,\cite{ding_new_2008, yeh_gadolinium-loaded_2007} and embedded with BCF-92 wavelength shifting (WLS) fibers (1.5\,mm diameter, Luxium Solutions\,\cite{BCF-92} ). To ensure the stability of photon collection on the fiber surface, each WLS fiber was wiped with a cleaning cloth soaked in alcohol before being fixed.

\begin{figure}[!ht]
    \centering
    \includegraphics[width=0.65\linewidth]{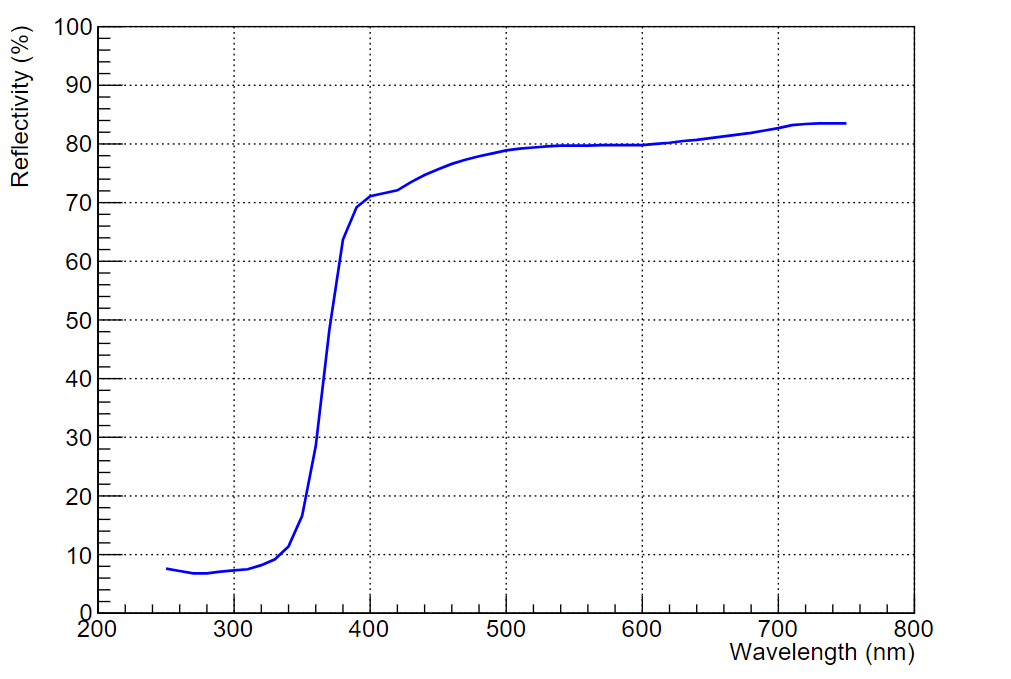}
    \caption{PVC reflectivity}
    \label{fig:PVC reflectivity}
\end{figure}

Due to compatibility issues between the LS and Tyvek~\cite{tyvek}, Tyvek was not used as the reflective material for the inner wall of the PVC. Instead, the reflection of photons relied on the inherent reflective properties of the PVC itself. 
The reflectivity of PVC was measured in air by the International Institute of Metrology, China\,\cite{https://www.nim.ac.cn/}. The results are shown in the Figure~\ref{fig:PVC reflectivity}.
A comparison will be made between the experimental test and the Monte Carlo simulation results.

\begin{figure}[htbp]
\centering
\subfigure[PMT sleeve units constrution]{\includegraphics[width=0.32\textwidth]{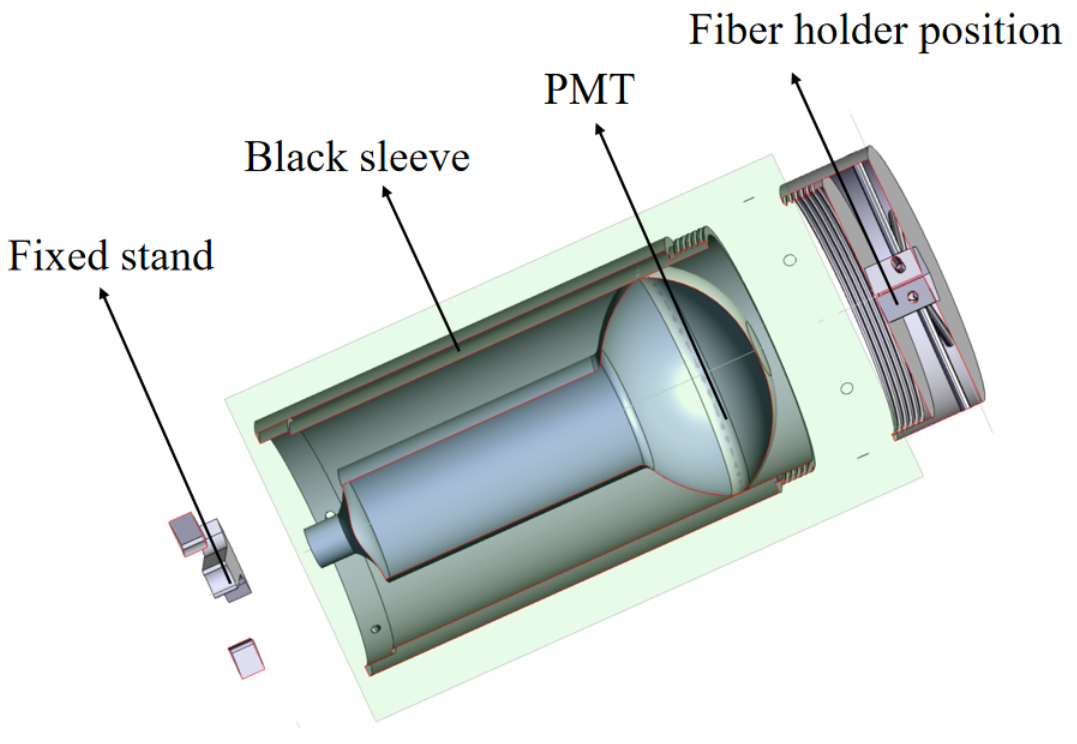}}
\subfigure[Fiber end face coupling with PMT]{\includegraphics[width=0.32\textwidth]{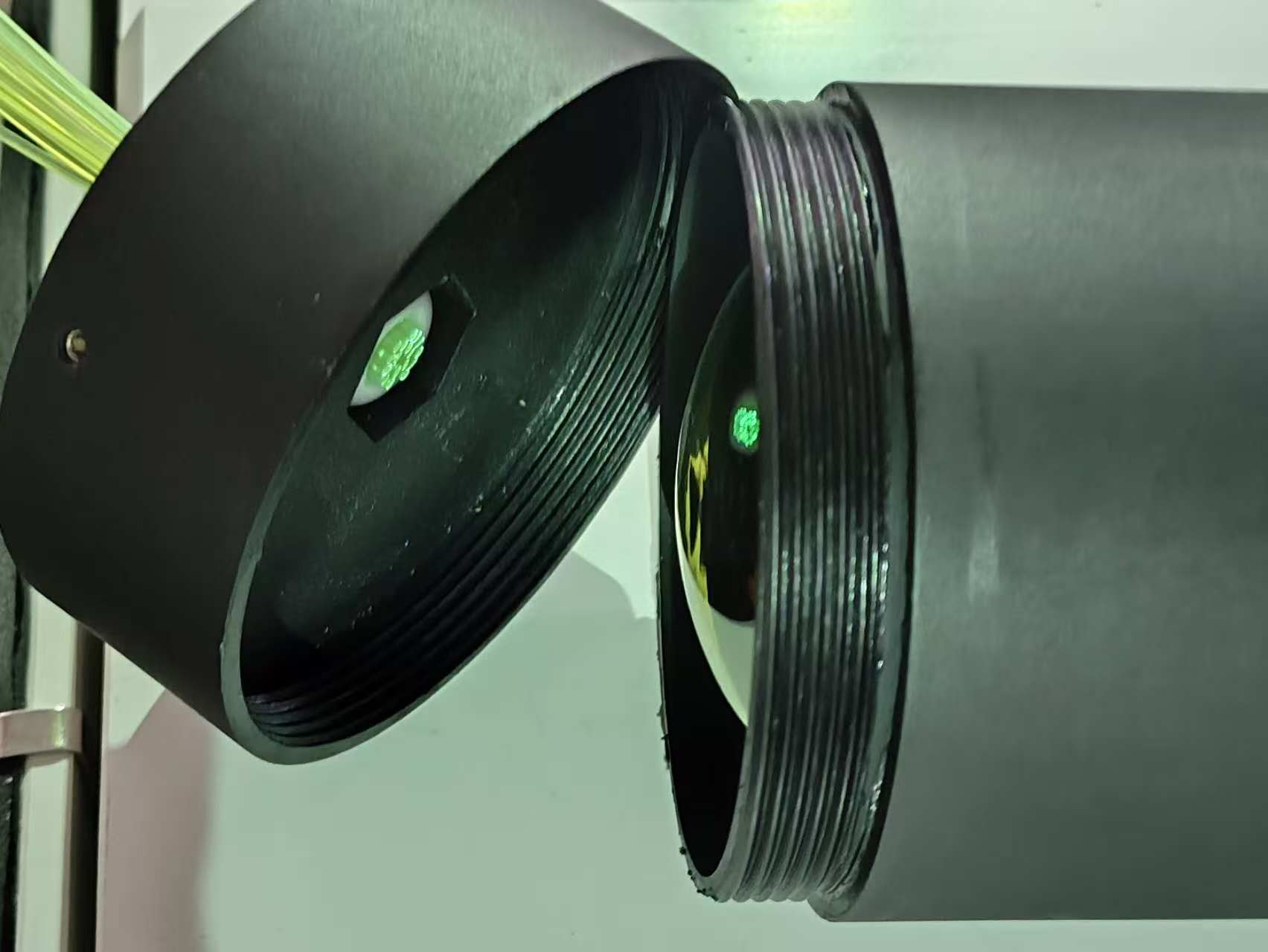}}
\subfigure[Fiber holder]{\includegraphics[width=0.28\textwidth]{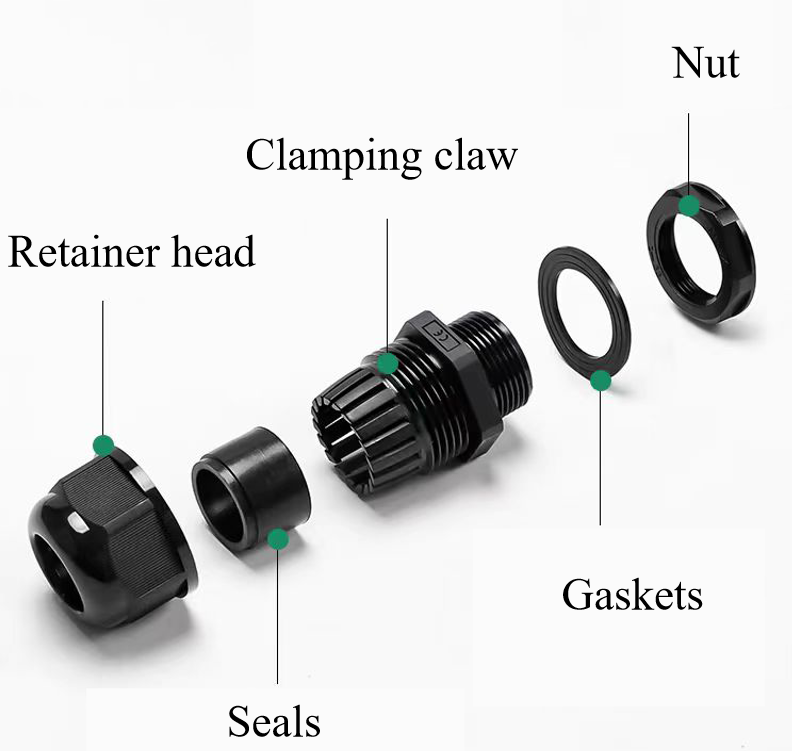}}
\caption{PMT sleeve units, WLS fiber secured and coupled with PMT photocathode.\label{fig:PMT sleeve units}}
\end{figure}

Thirty-two WLS fibers (Figure~\ref{fig:prototype module}), each 2.4\,m in length, are arranged in orthogonal layers: 16 horizontal and 16 vertical fibers.
The horizontal fibers are positioned 10\,mm above the container's bottom surface, with a uniform spacing of 40\,mm between adjacent fibers. The vertical fibers are stacked atop the horizontal layer at a height of 11.5\,mm.
All fibers are secured in place by pre-designed perforated PVC holding panels (3\,mm thick and 13.5\,mm high, as shown in Figure~\ref{fig:prototype module}c), which are welded to the bottom surface of the PVC container.
The fibers on each side are threaded through the holes at the four corners of the container (as shown in Figure~\ref{fig:prototype module}d). They are then screwed together using the fiber holders depicted in Figure~\ref{fig:PMT sleeve units}c. The coupling method between the fiber end and the PMT photocathode is shown in Figure~\ref{fig:PMT sleeve units}b. Next, the fiber holders are positioned in their corresponding locations, as illustrated in Figure~\ref{fig:PMT sleeve units}a. Finally, the fiber holders are secured in place with three screws.

Prior to PMT coupling, the ends of the fibers undergo a four-stage polishing protocol to improve the photon collection ratio: D30 (coarse), D9 (medium), D1 (fine), and XF5D (ultra-fine) to achieve optimal light transmission.
Fiber holders are used to secure the WLS fibers, ensuring that their ends are aligned on the same horizontal plane.
Silicone oil was applied to the fiber end faces to improve the coupling between the WLS fibers and the PMT photocathode window.
This setup increased the number of collected PE by almost 8\% \cite{Zhang:2020eyg,wang_testing_2021}.
Subsequently, the fiber ends are positioned on the photocathode surface of the PMT to minimize photoelectron loss.

Additionally, the exterior of each PMT sleeve is wrapped with an alloy that serves as magnetic shielding.
The signal readout and power cables are routed out of the aluminum box, and the entire prototype module is made light-tight. The gap between the lid and the box is sealed with a rubber sealing strip.

\begin{figure}[htbp]
\centering
\includegraphics[width=.4\textwidth]{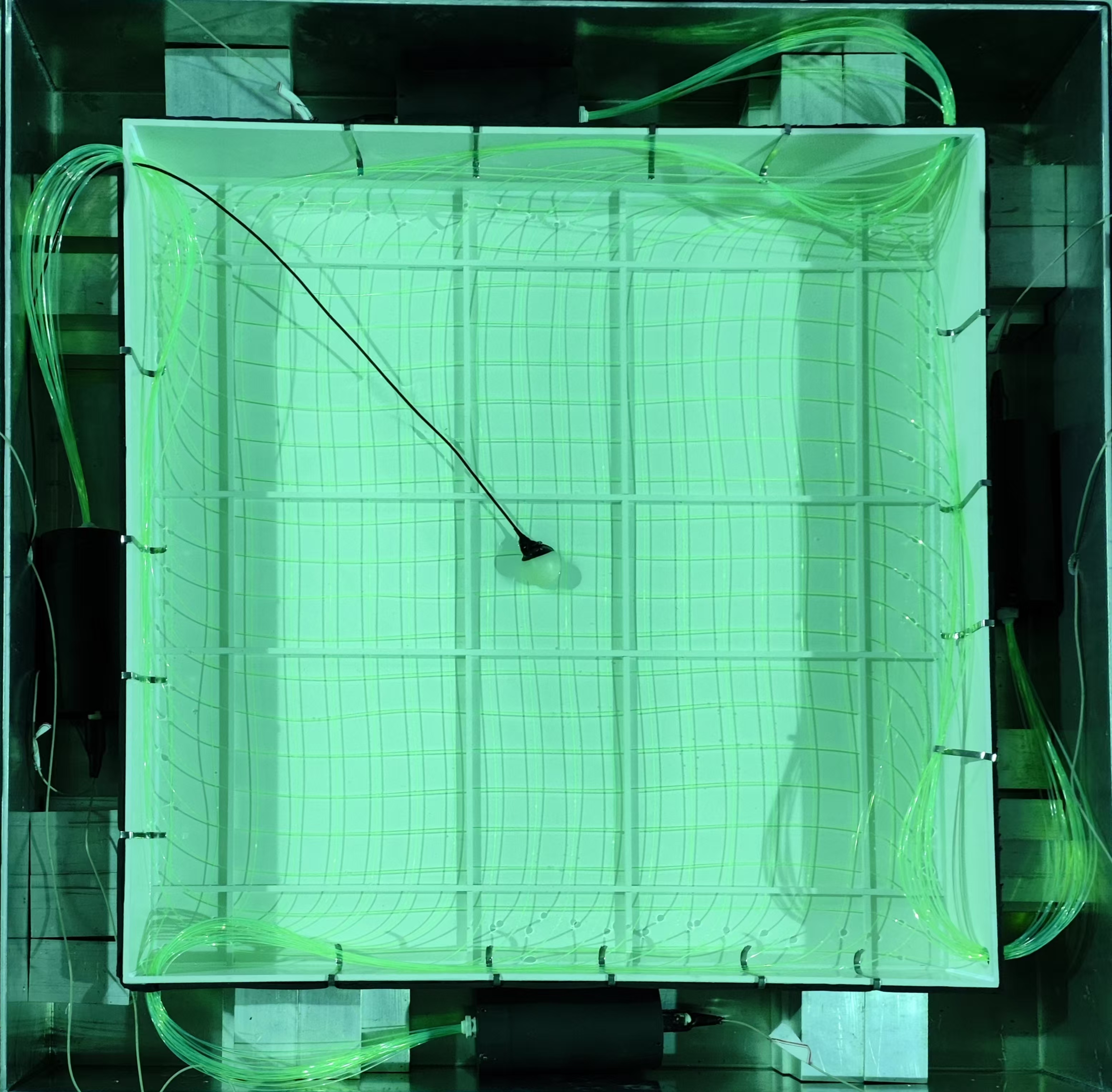}
\caption{Bird view of a liquid scintillation experiment.\label{fig:exp top view}}
\end{figure}

The realized prototype of the liquid scintillator module with WLS fiber is illustrated in Figure~\ref{fig:exp top view}. 
The LS is contained in the PVC container, with a WLS fiber positioned inside the container. Both ends of the WLS fiber are connected to 3-inch photomultiplier tubes (PMTs). 
Ionizing radiation passing through the LS generates optical photons, which are collected by the fibers and detected at both ends by the PMTs. 
The arrangement of multiple fibers in the longitudinal and transverse directions gives the detector a two-dimensional position tracking capability. Compared to the Liquid Scintillator with Fibers (LSF) detector\,\cite{Zhangyongpeng_2017,Zhangying_2025} with only a single directional fiber, this LS detector module collects more photoelectrons.

\subsection{3-inch PMT \& readout}

26000 3-inch photomultiplier tubes (SPMTs)\,\cite{Cao_2021} have been selected for JUNO, manufactured by Hainan Zhanchuang Photonics Technology Co., Ltd (HZC) in China. Four PMTs used in this experiment were re-used from the JUNO prototype experiments~\cite{Wang_prototype_2017}, where also using the PMTs of the same type.

\begin{figure}[!htbp]
\centering
\includegraphics[width=0.8\textwidth]{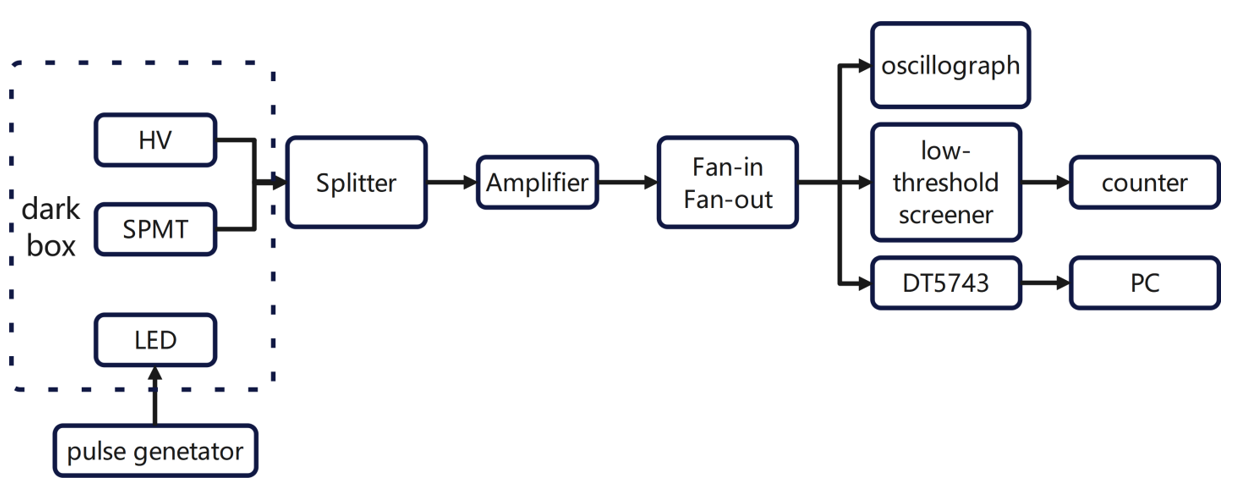}
\caption{PMT calibration logic flowchart.\label{fig:PMT calibration logic flowchart}}
\end{figure}

\begin{figure}[!ht]
\centering
\subfigure[SPE charge VS. HV]{\includegraphics[width=0.6\textwidth]{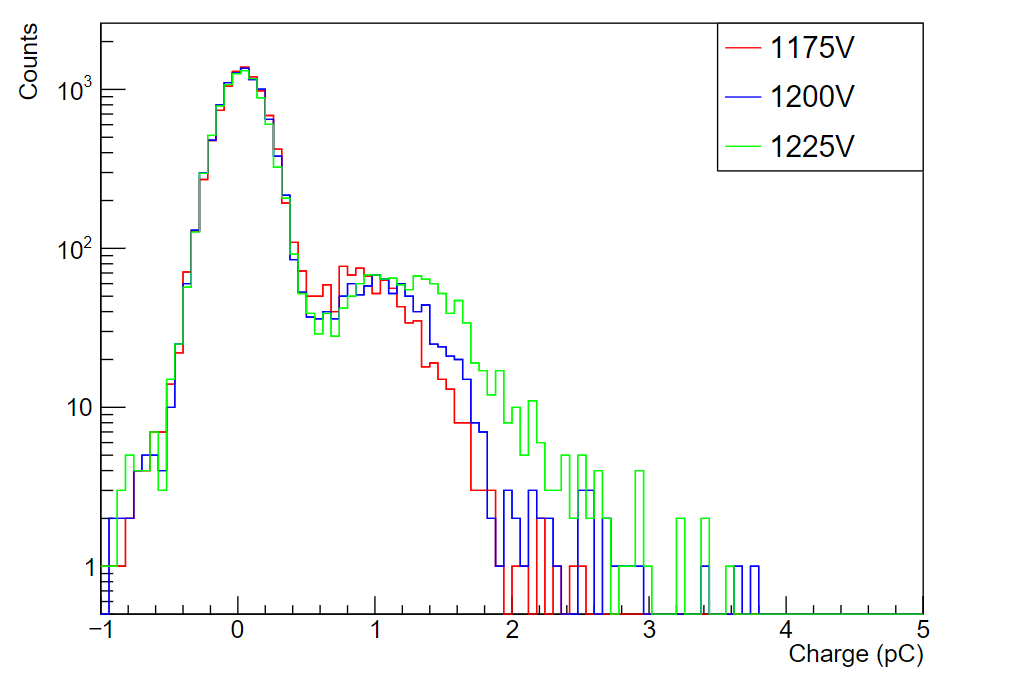}}
\subfigure[Gain VS. HV]{\includegraphics[width=0.45\textwidth]{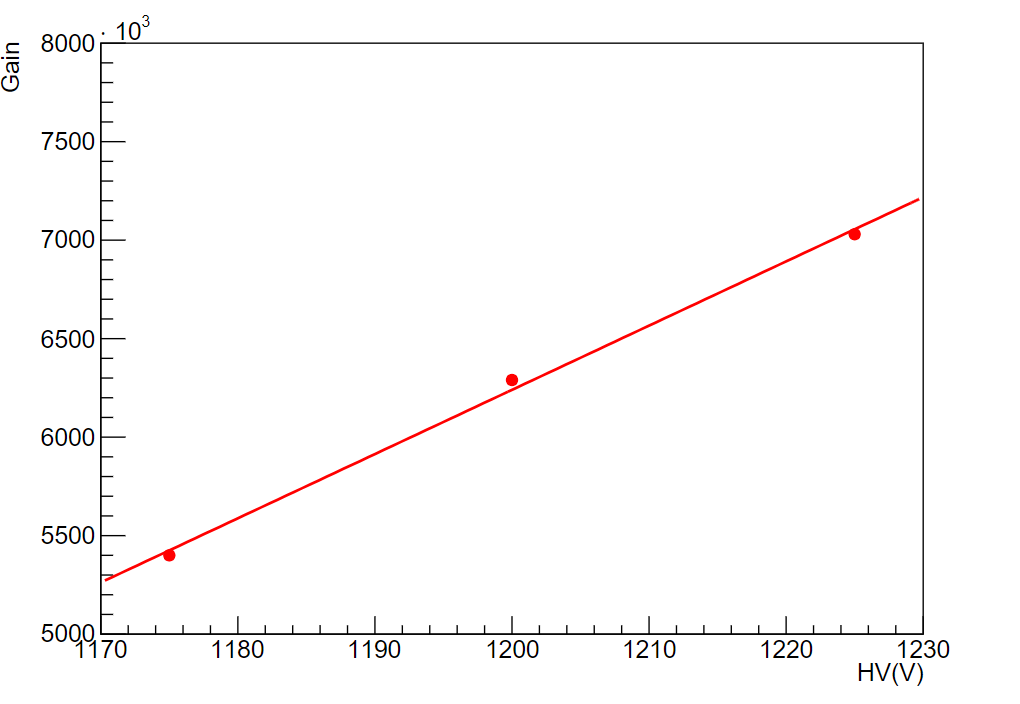}}
\subfigure[Rise-time, fall-time, and FWHM of 4 PMTs.]{\includegraphics[width=0.45\textwidth]{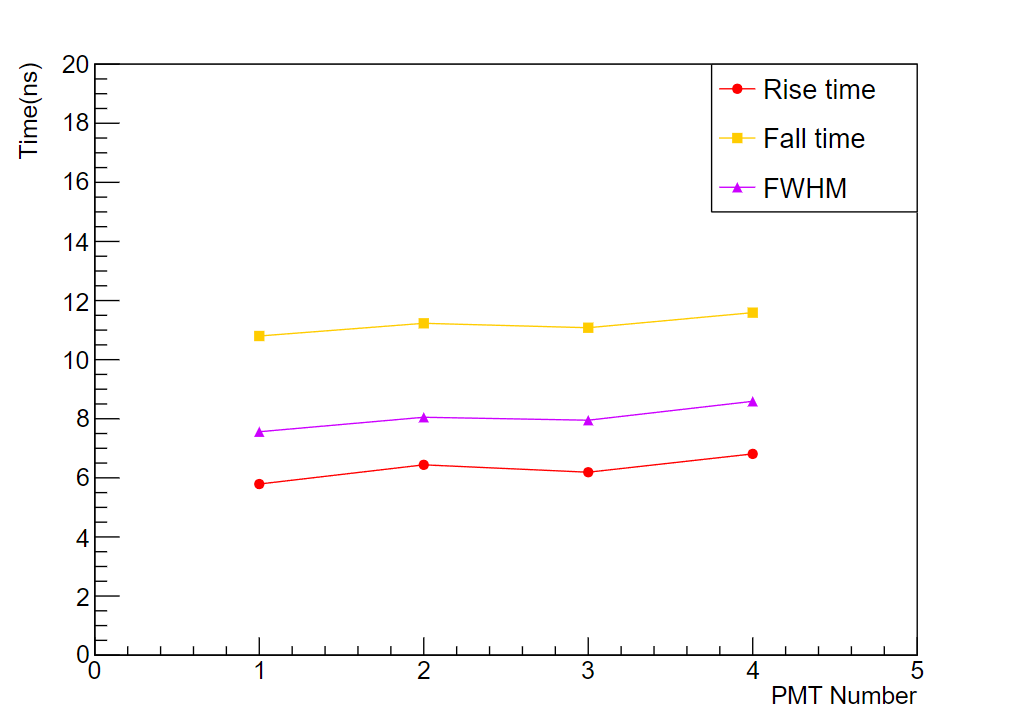}}
\caption{PMT testing}
\label{fig:SPMT testing}
\end{figure}

The four 3-inch PMTs (HZC XP72B22)\cite{20190106,wu_study_2022,Cao_2021} are individually measured in a dark box prior to installation using a test system with LED illumination at the single-photon electron (SPE) level. Data acquisition is performed using a CAEN DT5743\,\cite{DT5743} digitizer.
The DT5743 is the core component of the test system (Figure~\ref{fig:PMT calibration logic flowchart}). It captures the PMT output waveforms to calculates its charge, and determines the gain versus high voltage (HV) by measuring voltages, as shown in Figure~\ref{fig:SPMT testing}.

To maintain consistency on the threshold impact, the gain of all four PMTs was adjusted to approximately $6\times10^6$ (around 4\,mV per SPE) through single-photon electron testing and LED illumination. The nominal high voltage for each PMT is shown in Table~\ref{tab:PMT High Voltage}.

\begin{table}[!htbp]
\centering
\caption{PMT High Voltage with the Gain of $6\times10^6$.\label{tab:PMT High Voltage}}
\smallskip
\begin{tabular}{c|c}
\hline
PMT Number&HV/V\\
\hline
1 & 1405 \\
2 & 1260 \\
3 & 1202 \\
4 & 1265 \\
\hline
\end{tabular}
\end{table}

Simultaneously, the rise-time, fall-time, and FWHM of the four PMTs were measured, as shown in Figure~\ref{fig:SPMT testing}c. The average rise-time is approximately 6\,ns, the average fall-time is around 11\,ns, and the average FWHM is about 8\,ns.


\section{Simulation}
\label{sec:sim}

\subsection{Setup}
\label{sec:Setup}

To achieve an enhanced signal-to-noise ratio, the detector is trying to design to collect a higher number of PE, allowing for an appropriate discrimination threshold to be set.
Factors influencing PE collection include the thickness of the LS, the LS attenuation length, container reflectivity, PMT quantum efficiency, WLS fiber attenuation length, the nature of the couplings between WLS fibers and PMTs, and other factors.
The design of the 80\,cm × 80\,cm× 12\,cm PVC container ensures that photons travel only tens of centimeters after reflecting several times.
According to a previous study\,\cite{ding_new_2008}, the LS attenuation length exceeds 10 meters. 
Therefore, the influence of the LS attenuation length can be considered negligible.
Most photons are absorbed on the walls of the LS container.

A Monte Carlo simulation of the module based on Geant4\,\cite{Geant4} is following the dimensions of the experimental setup.
All parameter settings related to the properties of the LS are derived from previous study \cite{JUNO-LS-ABUSLEME2021164823}.
The WLS fiber, model BCF-92, has a length of 2.4 meters and consists of a polystyrene-based core and a polymethylmethacrylate (PMMA) cladding. Its parameters are listed in Table~\ref{tab:BCF-92} \cite{BCF-92}.

\begin{table}[!htbp]
\centering
\caption{BCF-92 Fiber Parameters.\label{tab:BCF-92}}
\smallskip
\begin{tabular}{c|c}
\hline
Emission peak& 492\,nm\\
Decay time & 2.7ns\\
Attenuation Length& >4\,m\\
Core material&Polystyrene\\
Core refractive index & 1.60 \\
Core Density & 1.05\,g/cm$^3$\\
Cladding refractive index & 1.49 \\
Cladding thickness,round fibers & 3\% of fiber diameter\\
Cladding material&PMMA\\
Cladding Density & 1.2\,g/cm$^3$ \\
\hline
\end{tabular}
\end{table}

\begin{figure}[!htbp]
\centering
\subfigure[Wavelength shifter]{\includegraphics[width=0.32\textwidth]{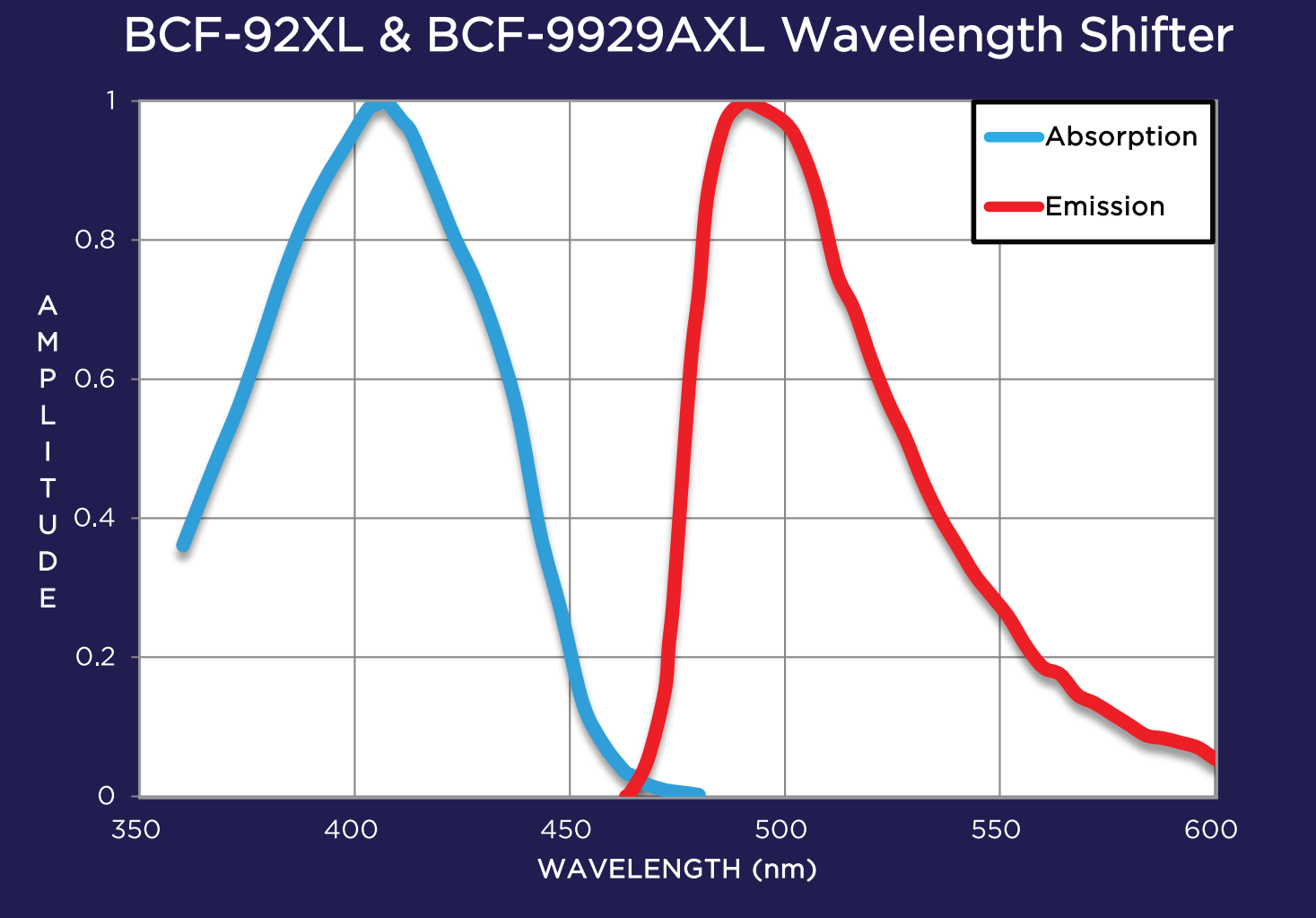}}
\subfigure[QE of 3-inch PMT]{\includegraphics[width=0.49\textwidth]{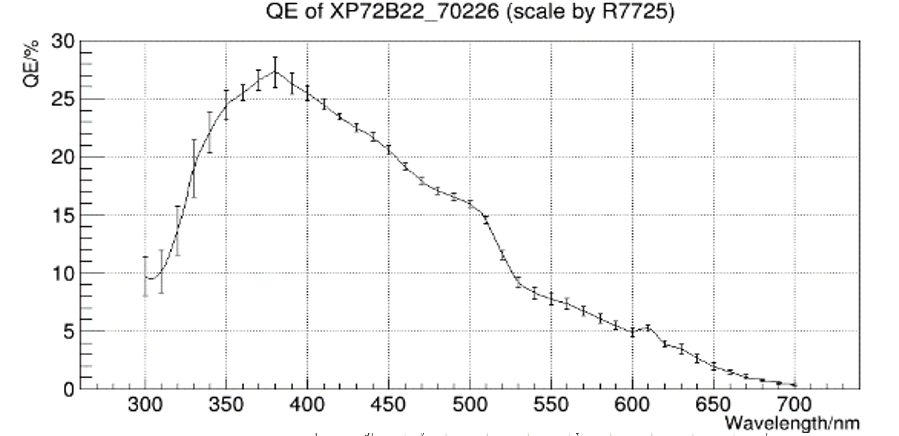}}
\caption{Wavelength of BCF-92 and PMT QE.\label{fig:wavelength}}
\end{figure}

The wavelength shifting of the fiber is shown in Figure~\ref{fig:wavelength}a \cite{Zhang:2020eyg}. The fiber in the simulation is placed without any bending and emerges from the container wall, and is not secured using a fiber holding panel. 
At the fiber end, four PMTs \cite{CAO2021165347} are placed and attached, and the quantum efficiency is set according to the measurements, as shown in Figure~\ref{fig:wavelength}b.


\subsection{Event simulation}
\label{sec:EvtSim}

\begin{figure}[!htbp]
\centering
\subfigure[Deposited energy of Muon (LS 8\,cm)]{\includegraphics[width=0.45\textwidth]{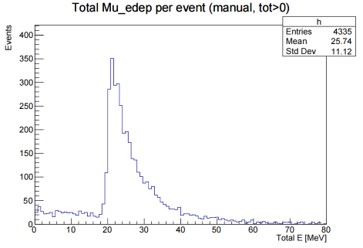}}
\subfigure[Photoelectron response to Muon and radioactive gamma (LS 8\,cm)]{\includegraphics[width=0.43\textwidth]{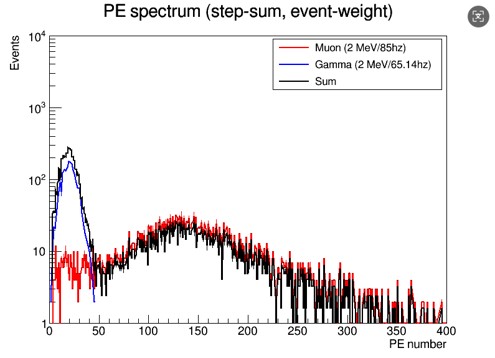}}
\caption{Simulated detector (LS 8\,cm).\label{fig:sim-muon}}
\end{figure}

Cosmic ray muons pass through the LS, depositing energy that produces scintillation photons. These photons are collected by the fibers and transmitted to the PMTs.
These photons are collected by WLS fibers and transmitted to the photocathode of the PMT.
The PMT then processes these signals through a series of steps to generate a measurable output signal.
The energy deposited by the muon follows a Landau distribution, while the number of scintillation photons produced by LS follows a Gaussian distribution. The simulated spectrum with LS thickness of 8\,cm are shown on Figure~\ref{fig:sim-muon}.

\begin{figure}[!htbp]
\centering
\subfigure[charge in PE]{\includegraphics[width=0.48\textwidth]{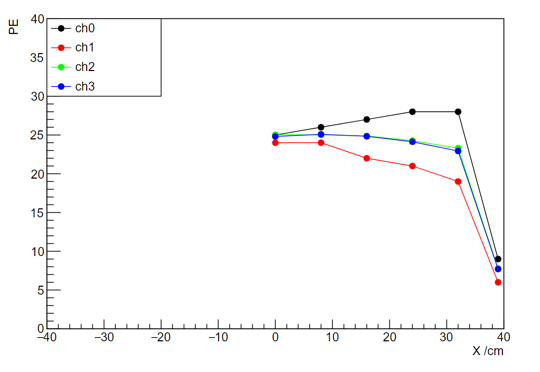}}
\subfigure[First-hit time.]{\includegraphics[width=0.48\textwidth]{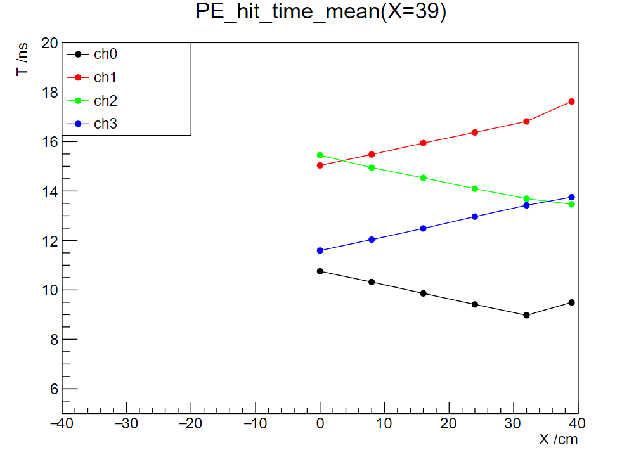}}
\caption{Detector Uniformity.\label{fig:Detector Uniformity}}
\end{figure}

To evaluate the detector's reconstruction performance, we initially conducted simulation tests to assess its time uniformity and PE quantity uniformity (Fig.~\ref{fig:Detector Uniformity}).
By examining muons at various incidence positions, we investigated the distribution of PE first-hit times on the PMTs and the variations in PE counts among the four PMTs.
Larger deviations in either the PE distribution or the time distribution enhance the position reconstruction capabilities, thereby improving the detector's vertex resolution.
The incident muon simulated firstly is perpendicular to the detector surface and has an energy of 4\,GeV. The height Z of the incident point is 1\,m from the center plane of the LS, while the X and Y directions are distributed within the range of -400\,mm to 400\,mm.
By surveying the incidence position, we processed the data from the PMTs to obtain the PE count and their first-hit time of PEs arriving at the PMTs for each muon.
As observed from Figure~\ref{fig:Detector Uniformity}, the uniformity of the first-hit time shows significant variations across PMTs as the incidence position is varied in one direction. 
Consequently, the first-hit time data can be used to reconstruct the incident position of the muon.

By simulating the first-hit time of PEs arriving at the PMTs, the incident particle positions were reconstructed using a time-centered algorithm.

\section{Measurement}
\label{sec:test}

\subsection{Coincidence level}
\label{sec:test:Water}

\begin{figure}[htbp]
\centering
\includegraphics[width=.9\textwidth]{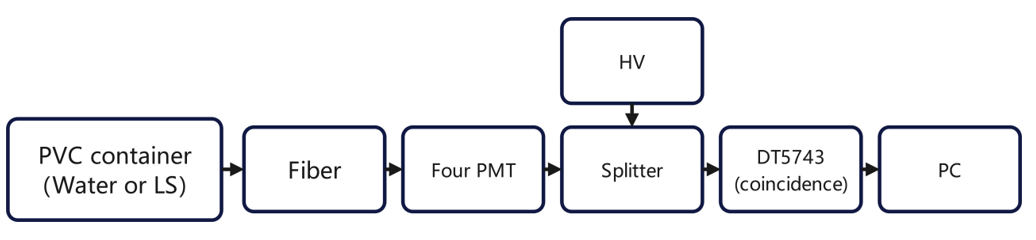}
\caption{Experimental flowchart.\label{fig:Experimental flowchart}}
\end{figure}

The output signals from the four PMTs were routed through a splitter to the DT5743, which was used for waveform acquisition.
Thresholds of 4\,mV or 6\,mV were set to collect signal waveforms under different coincidence levels/modes: 1/4 (1 out of 4), 2/4, 3/4, and 4/4 triggering configurations. Both the threshold settings and the coincidence modes were configured via the DT5743's software interface.
Figure~\ref{fig:Experimental flowchart} shows the basic logic diagram of the measurements.

Prior to liquid filling, the detector was tested in air firstly.
The four PMTs operate at the same gain of $6\times10^6$, and the high voltages (ranging 1202 to 1405\,V in Table~\ref{tab:PMT High Voltage}) applied to each PMT are different.
Their corresponding dark count rates (DCRs) also are different as shown in the Table~\ref{tab:DCR of each PMT}.
In the case of four-PMT coincidence (4/4), the rate in the air phase is observed to be very small, indicating that the noise has been effectively suppressed. 
The system exhibited a four-PMT coincidence (4/4) rate of 0.023\,Hz at a 4\,mV threshold within a 90\,ns time window, while the random coincidence rate by calculation should be less than 0.001\,Hz. This rate decreased to undetectable levels (<0.001 Hz) when the threshold was increased to 6\,mV. 

\begin{table}[!htbp]
\centering
\caption{DCR for each PMT under different thresholds.\label{tab:DCR of each PMT}}
\smallskip
\begin{tabular}{c|cc}
\hline
\multirow{2}{*}{\centering PMT Number} & \multicolumn{2}{c}{DCR/Hz} \\
\cline{2-3}
 & 4 mV & 6 mV \\
\hline
1 & 4958 & 497 \\
2 & 1147 & 132 \\
3 & 1726 & 332 \\
4 & 5754 & 259 \\
\hline
\end{tabular}
\end{table}

\subsection{Trigger rates}
\label{sec:trigger-rates}

\begin{figure}[!htbp]
\centering
\subfigure[Trigger rate of fired PMTs with Air, Water and different thicknesses]{\includegraphics[width=0.49\textwidth]{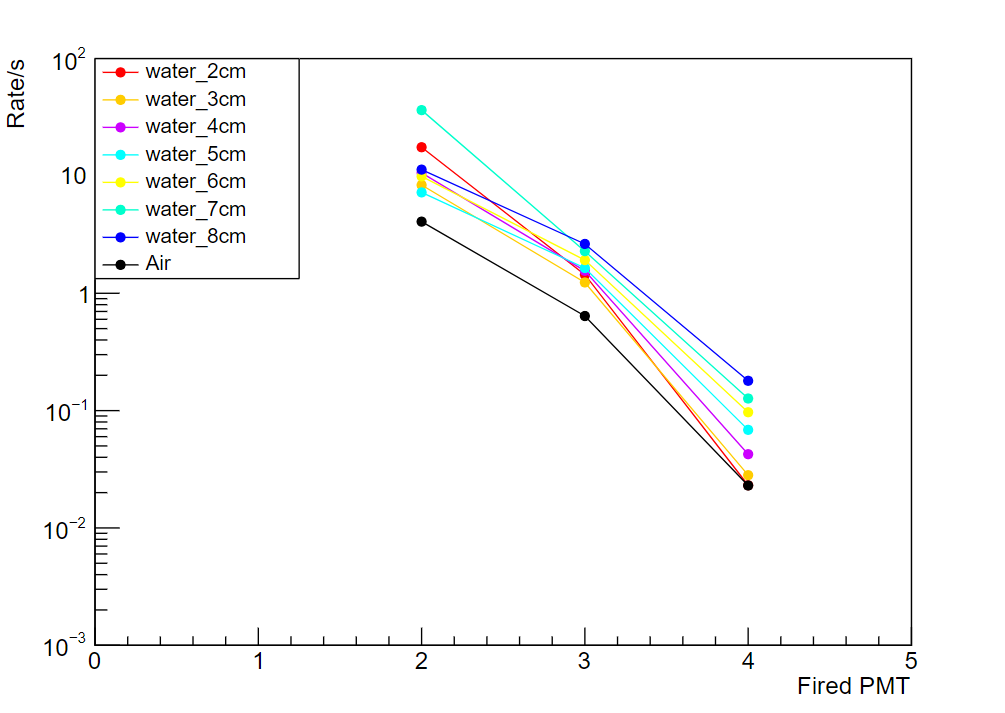}}
\subfigure[Trigger rate of fired PMTs with LS and different thicknesses]{\includegraphics[width=0.49\textwidth]{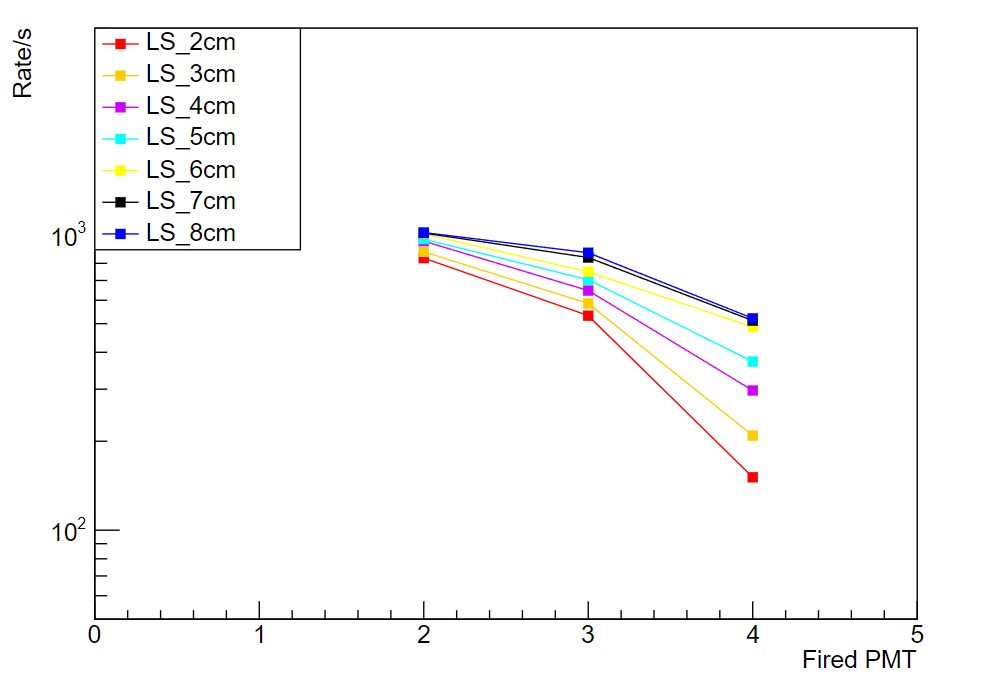}}
\caption{Trigger rates of the prototype under different coincidence level and different mediums.\label{fig:Rate}}
\end{figure}

Further more, the prototype is tested with pure water and LS at different liquid thicknesses and different trigger configurations as listed in Section \ref{sec:test:Water}.
Figure~\ref{fig:Rate}a and Figure~\ref{fig:Rate}b respectively show the rates in water and LS.
As the number of required fired PMTs increasing, the coincidence rate decreasing as expected.
At the same time, as the thickness of water or LS increasing, the coincidence event collection rate also increasing for all the trigger configuration.
If comparing to the rate under LS, the contribution from random coincidence under air (around 0.1\,Hz with 4/4 trigger mode) becomes negligible. The rate with LS and 4/4 trigger mode is affecting by radioactivity more when the LS thickness increasing, which will be further discussed in the following sections.

\section{Results \& comparison}
\label{sec:results}

\subsection{Detector response}
\label{sec:test:Dectector_response}

\begin{figure}[!htbp]
\centering
\includegraphics[width=.75\textwidth]{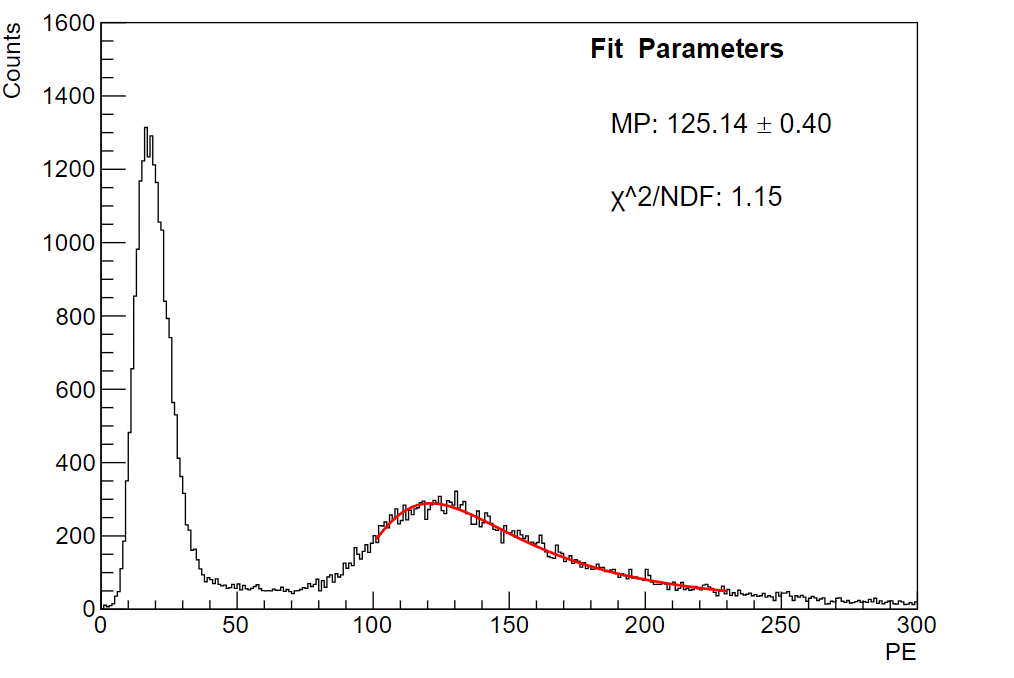}
\caption{Detector response with the 4/4 coincidence mode and an 8\,cm thickness of LS.}
\label{fig:Detector_response}
\end{figure}

Figure~\ref{fig:Detector_response} shows a typical charge spectrum summed of all the 4 PMTs from the prototype with the 4/4 coincidence mode and an 8\,cm thickness of LS. It shows two individual peaks: the right peak is response to muons, and the left peak is from natural radioactivity. The most probable intensity of a muon is around 125\,PE. And the muon can be clearly identified from the natural radioactivity.


\subsection{Influence of LS thickness}
\label{sec:test:LS}

\begin{figure}[!htbp]
\centering
\subfigure[PE distribution at different thickness of LS.]{\includegraphics[width=0.45\textwidth]{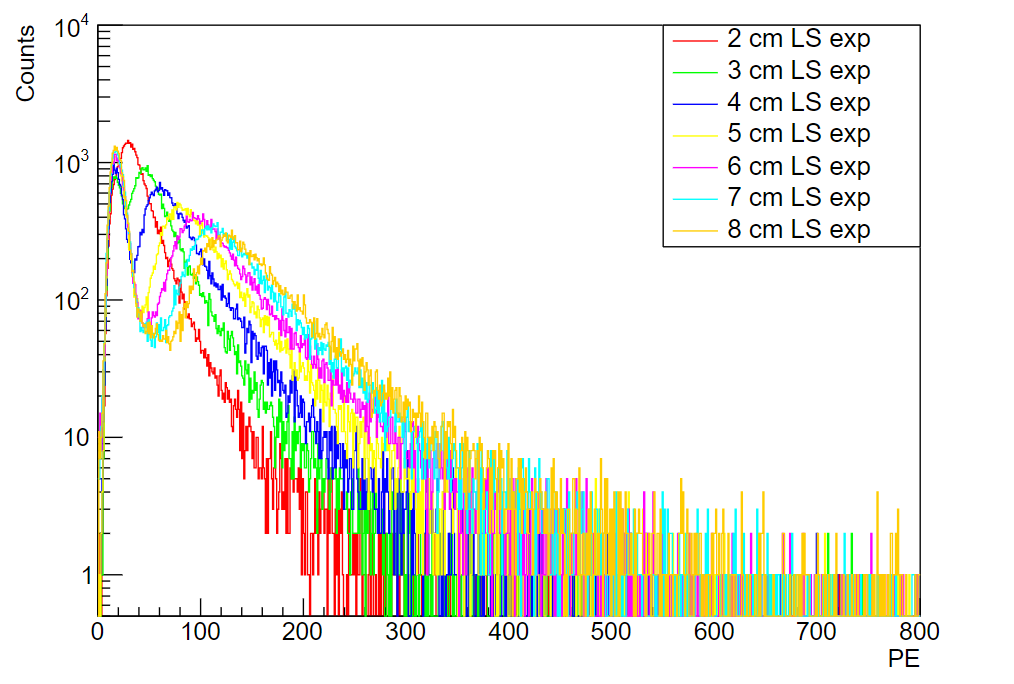}}
\subfigure[Experimental and Geant4 Monte Carlo comparison of the number of collected PEs.]{\includegraphics[width=0.45\textwidth]{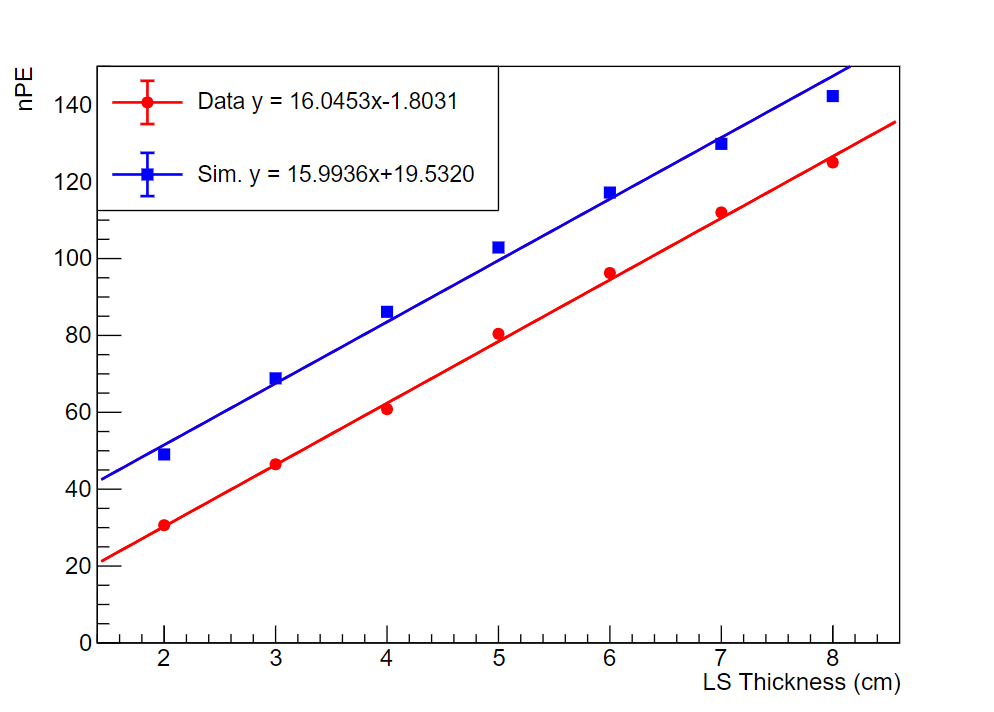}}
\caption{Light yield vs. LS thickness.\label{fig:nPE_vs.LS_thickness}}
\end{figure}

As shown in Figure~\ref{fig:nPE_vs.LS_thickness}a, when the thickness of the LS is 2\,cm, the collected PE spectrum exhibits only a single peak responding to muon.
However, when the LS thickness exceeds 3\,cm, the PE spectrum shows two distinct peaks. The left peak, located at the lower end, is attributed to natural radioactivity, while the right peak is caused by muons. 
As the thickness of the LS increases, the second peak shifts further to the right. As the thickness of the LS increases, the number of photoelectrons collected by the detector also increases, which is advantageous for distinguishing cosmic-ray muons from environmental radioactivity.
The relationship between LS thickness and the number of collected PE was investigated, as understanding this relationship is crucial for optimizing future detector designs.

As the function of LS thickness, the number of collected PE almost shows a linear relationship. 
Both the simulation data and experimental data are fitted with linear functions (Y=k*x+b), where Y is the measured light intensity, and x is the thickness of LS, as shown in Figure~\ref{fig:nPE_vs.LS_thickness}b. The k value indicates that the number of PE generated per centimeter thickness of the detector module increases by approximately 16\,PE/cm for each additional centimeter of LS. The k values from the experimental and simulated data are very consistent, while the intercepts differ by 11.3\,PE. The difference on the intercept between data and simulation is from the supporting structure of WLS fiber used in the real detector (as shown in Fig.\ref{fig:prototype module}c), which blocked the photons when the LS thickness around 1.1\,cm, which is not realized in the simulation.

\subsection{Muon rate}
\label{sec:test:Muon rate}

\begin{figure}[!htbp]
    \centering
    \subfigure[PE Distribution Threshold Setting.]{
        \includegraphics[width=0.48\textwidth]{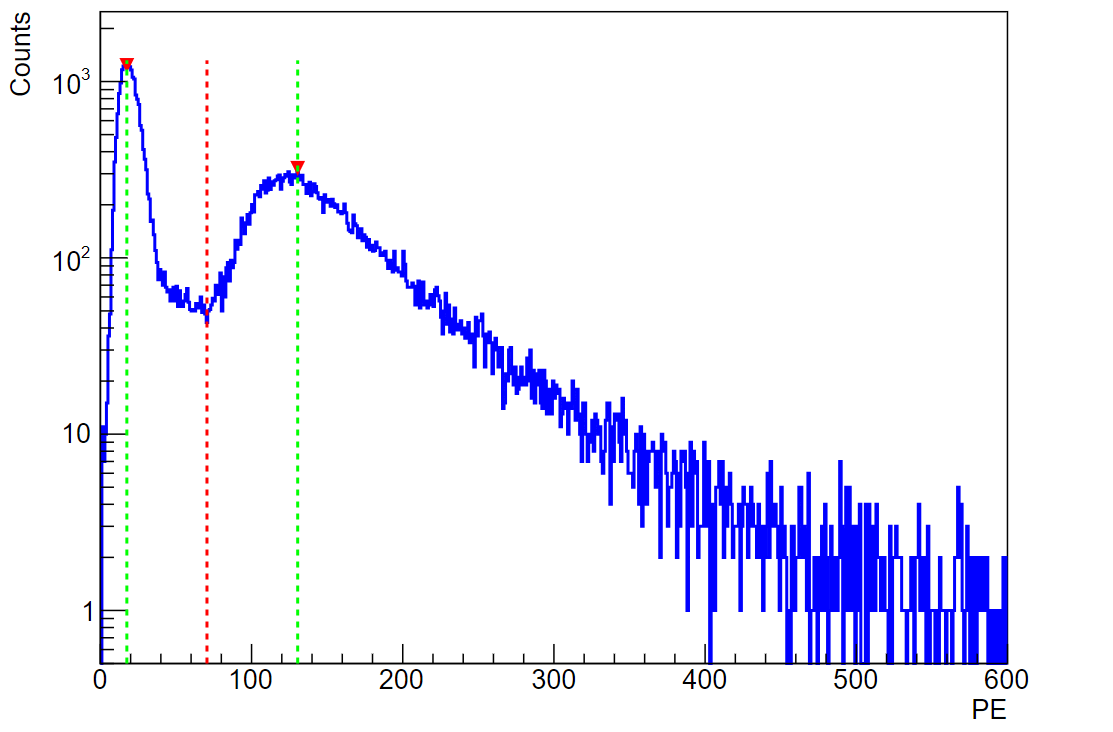}
    }
    \subfigure[Muon Rates for Different Coincidence Modes and Liquid Scintillator Thicknesses.]{
        \includegraphics[width=0.48\textwidth]{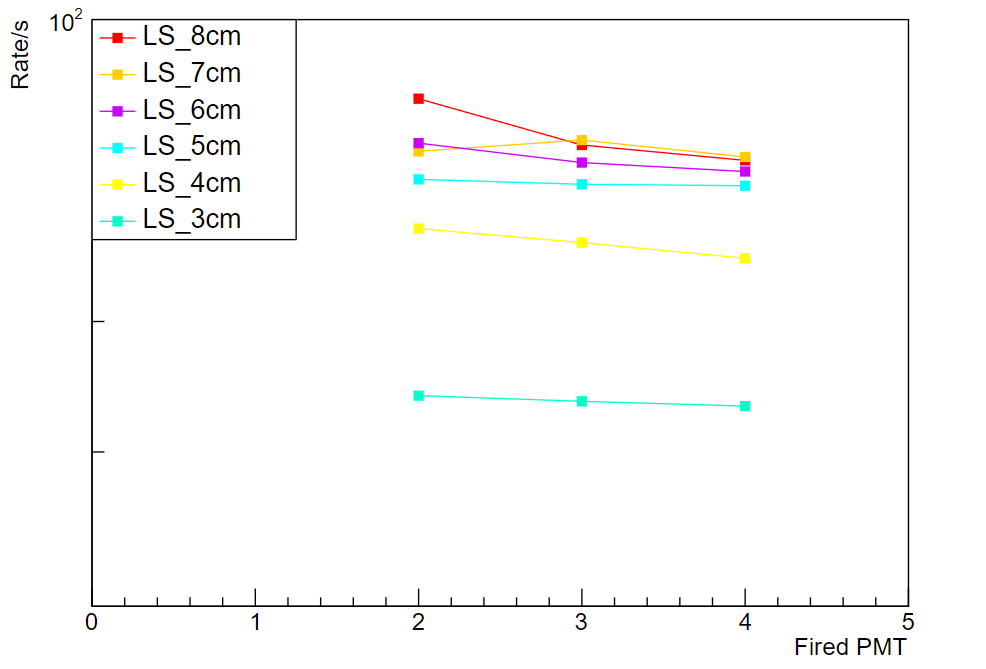}
    }
    \caption{Analysis of PE Threshold and Muon Rates.\label{fig:Muon Rates}}
\end{figure}

Based on the PE distribution of the two distinct peaks, the valley between these two peaks was selected as threshold to identity the natural radioactivity and muon. The events with PE less than this threshold were excluded, and the remaining events are considered as muons to calculate the muon rate.
The threshold for cutting is updating for different liquid scintillator thicknesses. For a 2\,cm thick liquid scintillator, it is not possible to set a cut. For a 3 cm thickness, the threshold cut is set at 28.5\,PE, while for an 8\,cm thickness, the threshold cut is at 70.5 PE. The red line in Figure~\ref{fig:Muon Rates}a illustrates the threshold cut for the 8\,cm case.

The muon rates for different coincidence modes and different liquid scintillator thicknesses are shown in Figure\ref{fig:Muon Rates}b.
From the overall data, it is clear that the identified muon rate increases with the thickness of the LS. The increase is particularly significant in the range of 3\,cm to 5\,cm, while the growth becomes less pronounced in the range of 6\,cm to 8\,cm. Beyond 5\,cm, the muon rate exceeds 80\,Hz and stabilizes around 85\,Hz for LS thicknesses between 6\,cm and 8\,cm. It means that the identification efficiency of muon is better with higher LS thickness. For an LS thickness of 8\,cm, the muon rate under 2/4 coincidence counting reaches approximately 91 Hz, corresponding to a rate of 84.7±0.8\,Hz under 4/4 coincidence counting, which is basically consistent to the expected ground muon flux and the detector dimension.

\subsection{Reconstruction}
\label{sec:Rec}

A muon typically generates signals to all the four PMTs. Each of the four PMTs has a first-arriving photoelectron, and their arrival times in the same direction (e.g., X direction) satisfy $ t_1 + t_2 = a $, where $ a $ is approximately constant. Therefore, the reconstructed position in the X direction can be expressed as:

\begin{equation}
\text{RecX} = x_1 \cdot \left(1 - \frac{t_1}{a}\right) + x_2 \cdot \left(1 - \frac{t_2}{a}\right),
\end{equation}

where $ x_1 $ and $ x_2 $ are the positions of the PMTs relative to the center of the detector, with values of $-1200 \, \text{mm}$ and $ 1200 \, \text{mm} $, respectively. The closer the muon is to a PMT, the smaller the time $ t $ as shown in Fig.\ref{fig:Detector Uniformity}. The position is initially corrected using the weight $ (1 - \text{weight}) $, where $ \text{weight} = \frac{t_1}{a} \,or\, \frac{t_2}{a}$. Once the reconstructed position is obtained, it is compared with the actual position to further correct the reconstruction. The same algorithm can also be applied to obtain the reconstructed position in the Y direction.
Here, the reconstructed position is rotated clockwise by $ 45^\circ $. To correct for this, the following transformation is applied:
\begin{align}
\text{RecX}^* &= (\text{RecX} - \text{RecY}), \\
\text{RecY}^* &= (\text{RecX} + \text{RecY}).
\end{align}

\begin{figure}[!htbp]
\centering
\subfigure[Distance to rec position X of simulation.]{\includegraphics[width=0.48\textwidth]{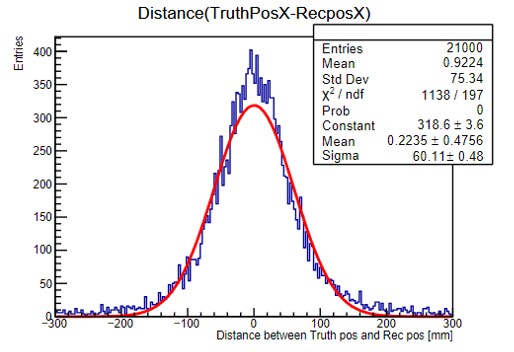}}
\subfigure[Reconstructed vertex for simulation: yellow points are the true position.]{\includegraphics[width=0.48\textwidth]{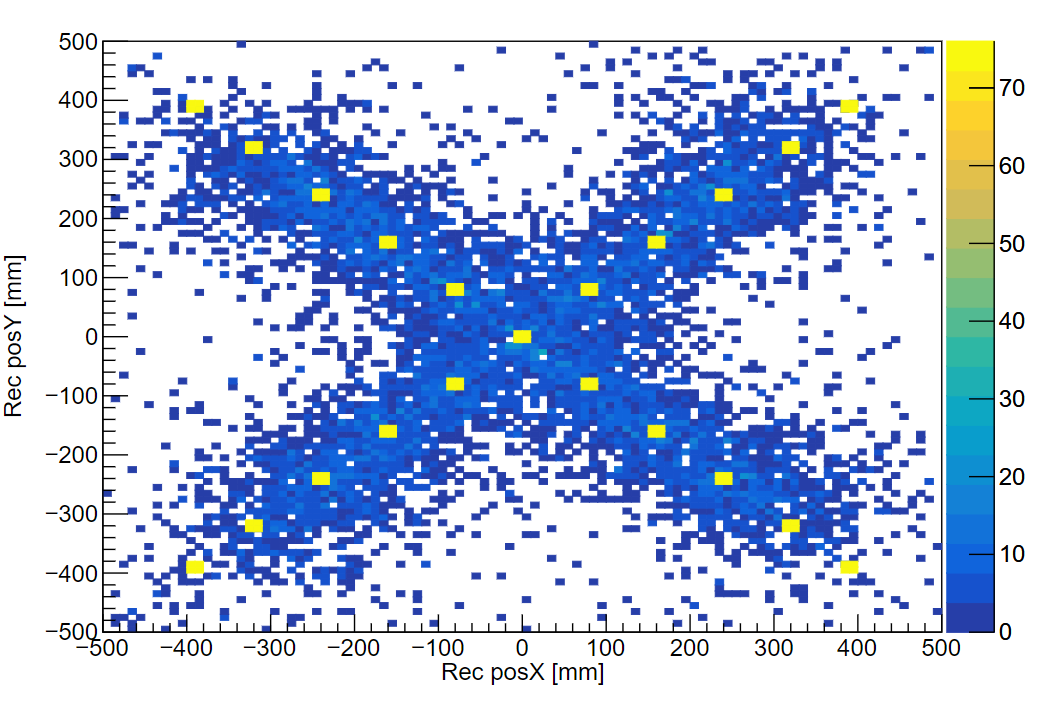}}
\subfigure[Reconstructed X for data.]{\includegraphics[width=0.48\textwidth]{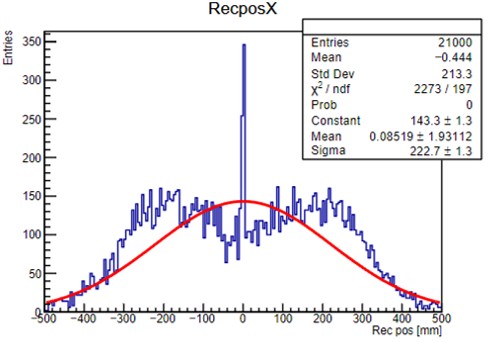}}
\subfigure[Reconstructed vertex for data.]{\includegraphics[width=0.45\textwidth]{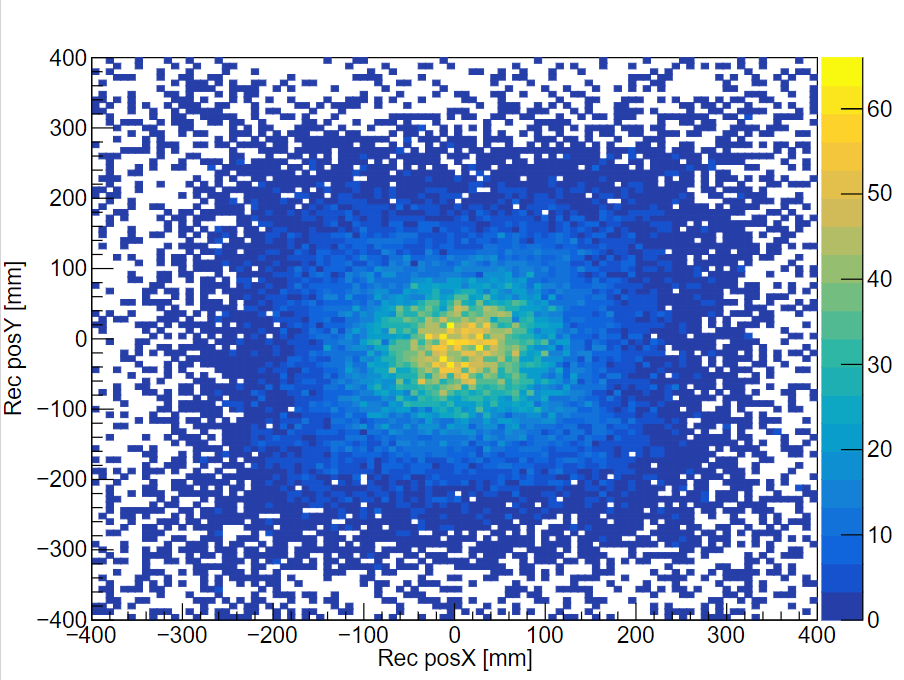}}
\caption{Vertex Reconstruction of data and simulation.\label{fig:rec-vertex}}
\end{figure}

A strong correlated between true and reconstruction position can be identified on the reconstructed 2D map for the simulation data with specified muon hit positions, as shown in Figure~\ref{fig:rec-vertex}b. And the distance between the reconstructed position and the true position is further checked, where a resolution of around 6\,cm on both X and Y axis is reached as shown in Figure~\ref{fig:rec-vertex}a. 

But for the data, the reconstructed vertex shows a centralized distribution to center for uniform muon samples as shown in Figure~\ref{fig:rec-vertex}d. Because here we can't identify the real hitting location of muon, only the reconstructed position is plotted on Figure~\ref{fig:rec-vertex}c, where also shows a similar centralization to the center except a peak around zero. It shows similar result on Y axis. The algorithm on data is not good as on simulation, which still needs a further investigation.


\section{Summary}
\label{sec:sum}

A detector prototype with wavelength shift fibers and liquid scintillator has been designed and built. 
A detailed investigation on its performance was studied by muon measurement and MC for different thickness of LS.
Tests have demonstrated that this detector schema exhibits excellent cosmic muon identification capabilities, enabling the differentiation between environmental radioactivity and muon events.
A muon passing through the detector module generates an average of 125\,PE with 8\,cm LS thickness. It is around 16\,pe/cm of LS thickness. The measured muon rate of the module is around 85\,Hz, and a good separation between radioactivity and muon. A fast reconstruction algorithm on vertex is proposed according to its timing features, but it shows worse results for data rather than simulation. The long term stability of LS with fiber still needs further investigation. It is valuable for further comparing on the detector options with different designs.

\acknowledgments
This work was supported partially by the National Natural Science Foundation of China (Grant No. 12505207, 11875282), the Natural Science Foundation of Hunan Province (Contract Nos.2025JJ60063), the State Key Laboratory of Particle Detection and Electronics (E429T1TD1, SKLPDE-ZZ-202208), the National Key Research and Development Program of China (Project No.2022YFA1602001, 2024YFE0110503), the Strategic Priority Research Program of the Chinese Academy of Sciences (Grant No. XDA10011200), and the Youth Innovation Promotion Association of CAS.














\bibliographystyle{JHEP}
\bibliography{biblio}

\end{document}